\documentclass[conference]{IEEEtran}
\usepackage{amsmath, amssymb, bm, cite, epsfig, psfrag, mathtools}
\usepackage{epstopdf}
\usepackage{graphicx}
\usepackage{algorithm,algpseudocode}
\usepackage{array}
\usepackage[margin=10pt,font=small]{caption}
\usepackage{bbm}
\usepackage{multirow}
\usepackage{fancybox}
\usepackage[usenames,dvipsnames]{xcolor}

\usepackage{subfigure}
\usepackage{etoolbox}
\usepackage{pbox}

\usepackage{tikz}
\usetikzlibrary{calc}

\newtoggle{conference}
\graphicspath{{figures/}}

\def\beq{\begin{equation}}
\def\eeq{\end{equation}}
\def\beqa{\begin{eqnarray}}
\def\eeqa{\end{eqnarray}}
\def\beqan{\begin{eqnarray*}}
\def\eeqan{\end{eqnarray*}}

\def\PL{\mathrm{PL}}
\def\dB{\mathrm{dB}}

\def\tm1{t\! - \! 1}
\def\tp1{t\! + \! 1}

\begin{document}

\thispagestyle{empty}

\bibliographystyle{IEEEtran}
\title{73 GHz Wideband Millimeter-Wave Foliage and Ground Reflection Measurements and Models}

\author{
	Theodore S. Rappaport,
	Sijia Deng
\IEEEauthorblockA{\\ 
NYU WIRELESS\\NYU Polytechnic School of Engineering, Brooklyn, NY 11201\\ tsr@nyu.edu, sijia@nyu.edu 
}
}

\maketitle

\begin{tikzpicture} [remember picture, overlay]
\node at ($(current page.north) + (0,-0.25in)$) {T. S. Rappaport, S. Deng, ''73 GHz Wideband Millimeter-Wave Foliage and Ground Reflection Measurements and Models,''};
\node at ($(current page.north) + (0,-0.4in)$) {\textit{in 2015 IEEE International Conference on Communications (ICC), ICC Workshops}, 8-12 June, 2015.};
\end{tikzpicture}
\thispagestyle{empty}

\begin{abstract}

This paper presents 73 GHz wideband outdoor foliage and ground reflection measurements. Propagation measurements were made with a 400 Megachip-per-second sliding correlator channel sounder, with rotatable 27 dBi (7$^\circ$ half-power beamwidth) horn antennas at both the transmitter and receiver, to study foliage-induced scattering and de-polarization effects, to assist in developing future wireless systems that will use adaptive array antennas. Signal attenuation through foliage was measured to be 0.4 dB/m for both co- and cross-polarized antenna configurations. Measured ground reflection coefficients for dirt and gravel ranged from 0.02 to 0.34, for incident angles ranging from 60$^{\circ}$ to 81$^{\circ}$ (with respect to the normal incidence of the surface). These data are useful for link budget design and site-specific (ray-tracing) models for future millimeter-wave communication systems.

\end{abstract}


\section{Introduction}

The continual demand for user capacity in wireless communication systems has motivated the use of millimeter-wave (mmWave) spectrum, where a vast amount of raw bandwidth is available to support multi-gigabits-per-second (Gbps) data rates \cite{5GItwillwork,ZPiIntro2011,Ghosh2014JASC}. There is relatively little knowledge of mmWave propagation in mobile environments, information needed to create site-specific and statistical models to develop air interface standards, and to design and deploy future communication systems \cite{TedNewBook}. Vegetation is a common aspect of outdoor urban and suburban environments, and impacts the quality of a mmWave communication link due to its additional attenuation, scattering, and de-polarization effects \cite{FeinianModel,FernandesModel,weissberger1982,Dilworth38G}. Understanding and accounting for the foliage effcts on propagation will be vital for future mmWave radio-system design.

Previous work has already provided valuable insight into mmWave propagation through foliage, where directional antennas were used at both the base station and mobile terminals.  Papazian \textit{et al}. found that attenuation caused by trees at 28 GHz was 16-18 dB with no leaves present and 26-28 dB with leaves present \cite{28GSeasonal}. In \cite{Impact28G}, a vegetation-dependent attenuation factor was obtained by comparing the mean received power for line-of-sight (LOS) and non-LOS (NLOS) links, that ranged from 19 dB to 26 dB at 28 GHz, depending upon the polarization of the transmitted signal. Vogel's tree attenuation measurements at 20 GHz found median attenuation coefficients of 0.75 dB/m for bare pecan trees and 2.5 dB/m for pecan trees with leaves. Schwering's experiments conducted at 28.8 GHz and 57.6 GHz demonstrated that vegetation loss increased with frequency and foliage depth, at a rate of 1.3-2.0 dB/m for the first 30 m through the foliage and only 0.05 dB/m beyond 30 m \cite{mmWaveVegetation}. Attenuation measurements in \cite{Hao38GHz} showed 17 dB attenuation through dense canopy of an oak tree measured at 38 GHz. Measured data in \cite{Dilworth38G} indicated that the attenuation of 2-4 dB/m and 6-8 dB/m at 38 GHz could be attributed to dry and wet leaves, respectively. Additionally, ground reflection has been studied at both microwave and mmWave frequencies for the past few decades\cite{StraitonGroundRef,GroundRef}. The reflection coefficients measured in \cite{GroundRef} were 0.43 and 0.78 for ground (forest floor) at nadir incidence (normal incident angles of ${\pm}$1$^{\circ}$ and 0$^{\circ}$). 

In this paper, foliage attenuation models for wideband mmWave outdoor measurements at 73 GHz are presented to characterize mmWave propagation through vegetation, and showing attenuation (expressed as a factor \cite{vogel1999effects,goldhirsh1998handbook}) in excess of free space propagation. Finally, m-easured ground reflection coefficients are compared with the theoretical Fresnel reflection coefficients for different estimated soil permittivity values.
     
\section{Hardware System and Measurement Description}     

\subsection{Measurement Hardware}

The 73 GHz propagation measurements were conducted using a 400 Megachip-per-second (Mcps) sliding correlator channel sounder with a -7.9 dBm transmit power, using a pair of 27 dBi steerable directional horn antennas with 7$^{\circ}$ half-power beamwidth (HPBW) in both azimuth and elevation planes at both the transmitter (TX) and receiver (RX). The baseband pseudorandom noise (PN) sequence was created by an 11-bit linear feedback shift register with a length of 2047. At the TX side, the baseband PN sequence clocked at 400 MHz was first mixed with an intermediate frequency (IF) at 5.625 GHz, and then modulated with a local oscillator (LO) signal at 67.875 GHz, to generate a 73.5 ${\pm}$ 0.4 GHz spread spectrum RF signal with an 800 MHz first null-to-null bandwidth.

At the RX side, the received RF signal was first downconverted to IF, then demodulated into its 400 MHz baseband in-phase (\textit{I}) and quadrature (\textit{Q}) components, which were then cross-correlated with a PN sequence identical to the transmit PN sequence, but clocked at a slightly lower rate of 399.95 MHz, resulting in a time-dilated cross-correlation, and a slide factor of 8000 \cite{Tedbook,Cox}. The TX and RX channel sounder block diagrams and additional system specifications can be found in \cite{ShuaiPIMRC,TR003}.

\subsection{Measurement Environment}

\begin{figure}
\centering
\includegraphics[width=3.5in]{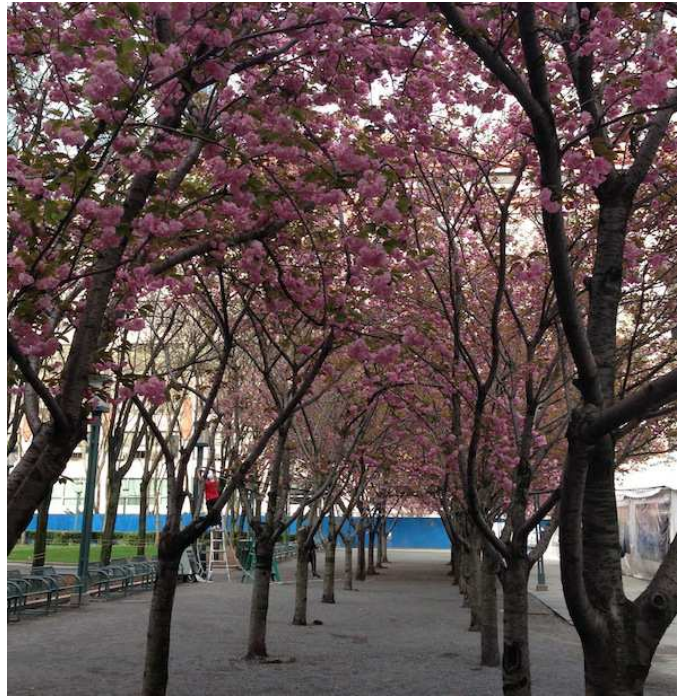}
\caption{Photograph of the MetroTech Commons Courtyard in downtown Brooklyn. The tops of the trees were approximately 6 m to 7 m AGL. The distance between two adjacent trees was approximately 4 m to 5 m. The ground consists of soil, dirt, gravel, and fallen leaves. The open area with concrete sidewalk is seen on the right of the stands of trees. }
\label{fig:Yard}
\end{figure}

\begin{figure}
\centering
\includegraphics[width=3.5in]{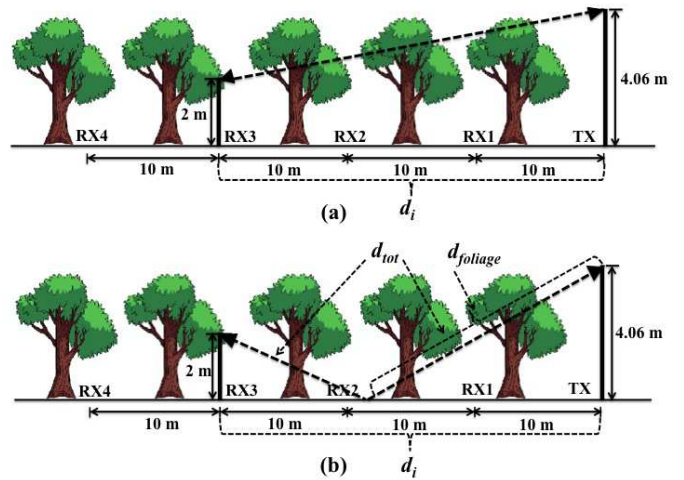}
\caption{Diagram of measurement scenarios for 73 GHz foliage and ground reflection measurements underneath the tree tops, where an example of the geometry at RX 3 is shown (note: measurements were made at all 4 RX locations). Fig.~\ref{fig:Foliage} (a) shows TX and RX locations with aligned antenna boresights for foliage penetration measurements. Fig.~\ref{fig:Foliage} (b) shows one of several ground reflection measurement configurations, where the TX and RX antennas are pointed at a common reflection point on the ground. Azimuthal sweeps were made for all antenna configurations.}
\label{fig:Foliage}
\end{figure}

The 73 GHz outdoor foliage and ground reflection measurements were conducted at New York University's MetroTech Commons Courtyard in downtown Brooklyn, a common courtyard-style area in a dense urban environment, shown in Fig.~\ref{fig:Yard}. The flat courtyard allowed measurements to first be made in an open sidewalk area, free from obstructions and far from neighboring buildings for free space calibration and testing (free space locations), and then permitted the equipment to be moved among a dense collection of cherry trees that lined the sidewalk area (foliage locations). T-R separation distances of 10 m, 20 m, 30 m, and 40 m were used for both the free space locations and the foliage locations. For all measurements, the TX antenna height was 4.06 m above ground level (AGL) to emulate lamppost height base stations, with the RX antenna heights set at 2.0 m AGL to emulate mobile users. The heights of the cherry trees were approximately 6-7 m, and the length of leaves varied between 7 and 14 cm. Measurements were made in late spring 2014, with moderately dense foliage on the trees, as seen in Fig.~\ref{fig:Yard}. A schematic of the measurement scenario for foliage measurements and ground reflection measurements is illustrated in Fig.~\ref{fig:Foliage}.

\subsection{Measurement Procedure}
\subsubsection{Free Space Measurements}

Free space measuremnts for co-polarized and cross-polarized antenna configurations were first conducted to determine the free space attenuation as reference for determine excess free space attenuation through the foliage. In the free space measurements, the TX and RX antennas were initially aligned boresight-to-boresight in both azimuth and elevation planes, the RX antenna was subsequently rotated in 10$^{\circ}$ increments in the azimuth plane, while the TX antenna remained fixed (boresight). A complete 360$^{\circ}$ RX azimuthal antenna sweep was performed at each RX location for both co-polarized vertical-to-vertical (V-V) and cross-polarized vertical-to-horizontal (V-H) antenna configurations. At each angular increment in the azimuth plane, a power delay profile (PDP) was acquired.

\subsubsection{Foliage Measurements}

 During the foliage measurements, six different unique pointing angle measurement sweeps were performed for each RX location underneath the tree canopy (see Fig.~\ref{fig:Foliage}), to investigate foliage-induced attenuation, scattering, and de-polarization effects. Using various elevation pointing angles at both the TX and RX, complete 360$^{\circ}$ sweeps were conducted at both the TX and RX antennas, from 10 to 40 m, for both V-V and V-H polarization configurations. The six sweeps included one AOA sweep where the TX and RX antennas were first perfectly aligned with each other on boresight in the elevation planes, while rotating the RX antenna over 360$^{\circ}$ in 10$^{\circ}$ increments in the azimuth plane. The one AOD sweep had the TX antenna rotated over 360$^{\circ}$ in 10$^{\circ}$ increments int the azimuth plane. At each azimuthal increment for the AOA and AOD sweeps, PDPs were recorded whenever energy was detected at the receiver for that unique pointing angle
combination. Then, four AOA sweeps were conducted for the case where the TX and RX antennas were uptilted and downtilted in the elevation plane, respectively, by 7$^{\circ}$ (one HPBW) with respect to the boresight elevation angle.

\subsubsection{Ground Reflection Measurements}

For the ground reflection measurements, the same TX and RX location combinations were used as in the foliage measurements. The ground consists of soil, dirt, gravel, and fallen leaves as shown in Fig.~\ref{fig:Yard}. At the beginning of the measurement, both the TX and RX antennas were initially downtilted by -30$^{\circ}$, -17$^{\circ}$, -11$^{\circ}$, -9$^{\circ}$ for the T-R separation distances from 10 to 40 m, respectively, such that the transmitted signal could be reflected off the ground into the RX antenna. The angles were calculated by assuming specular reflection off the ground. Five RX antenna sweeps were conducted, including one RX azimuthal sweep (Measurement 1, M1) with TX and RX elevation angles fixed at the initial angels for RX1 to RX4 with respect to horizontal, and four sweeps (M2-M5) for different TX and RX elevation angle combinations. Detailed TX and RX elevation combinations are shown in Table.~\ref{tbl:Ele}. For each measurement, the TX antenna was fixed and RX antenna was uptilted by 7$^{\circ}$ for each sweep, resulting in four different AOAs (\textit{e.g.} -30$^{\circ}$, -23$^{\circ}$, -16$^{\circ}$, -9$^{\circ}$ for RX1).  

\begin{table}
\caption{ TX and RX elevation angle combinations for all RX locations for the ground reflection measurements. Mea No denotes measurement number, and Ele denotes elevation. }
 \label{tbl:Ele}
 \centering
  \begin{tabular}{|c|c|c|c|}
	
	\hline
	 \textbf{RX ID} & \textbf{Mea No} & \textbf{TX Ele ($^{\circ}$)} & \textbf{RX Ele ($^{\circ}$)}  \\  \hline
	 \multirow{5}{*}{1}   &   1 &  -30   & -30    			    \\ \cline{2-4}
	                      &   2 &  -30   & -30,-23,-16,-9,-2    \\ \cline{2-4}
	 					  &   3 &  -23   & -30,-23,-16,-9,-2    \\ \cline{2-4}
						  &	  4 &  -16   & -30,-23,-16,-9,-2    \\ \cline{2-4}
						  &	  5 &  -9    & -30,-23,-16,-9       \\ \hline

	  \multirow{4}{*}{2}  &   1 &  -17   & -17    \\ \cline{2-4}
	 					  &   2 &  -17   & -17,-10,-3    \\ \cline{2-4}
						  &	  3 &  -10   & -17,-10,-3   \\ \cline{2-4}
						  &	  4 &  -6    & -17,-10,-3    \\ \hline
	  \multirow{3}{*}{3}  &   1 &  -11   & -11    \\ \cline{2-4}
	 					  &   2 &  -11   & -11,-4    \\ \cline{2-4}
						  &	  3 &  -11   & -11,-4   \\ \hline
	  \multirow{2}{*}{4}  &   1 &  -9    & -9    \\ \cline{2-4}
	 					  &   2 &  -9    & -9,-2   \\ \hline			 							  				 					  
     
 \end{tabular}
\end{table}

\section{Measurement Results and Analysis}

\subsection{Foliage Attenuation Path Loss Models}

Foliage attenuation path loss models were extracted by computing the omnidirectional total received power level at each RX location from the free space measurements, and comparing them with the omnidirectional power received from the foliage measurements. Omnidirectional path loss models were recreated by summing the powers received at each and every unique antenna pointing angle direction between the TX and RX. \cite{MKR:PIMRC14}. This is a valid approach, since signals from adjacent bins travelled different propagation distances, such that the phase of individual multipath can assumed to be random, thus allowing powers of each resolvable multipath component to be summed over the omnidirectional spatial manifold \cite{Tedbook}. The omnidirectional received powers and path losses are determined from the measurements using directional antennas, using the equations as follow \cite{MKR:PIMRC14}

\begin{equation}
\mathrm {Pr_{\mathrm{omni}}}[\mathrm{mW}] =  \sum_{z} \sum_{y} \sum_{x} \sum_{w} \mathrm{Pr}(\theta_{r_w},\phi_{r_x},\theta_{t_y},\phi_{t_z})
\end{equation}
\begin{equation}\label{E2}
\PL_{\mathrm{omni}}[\dB] = \mathrm{Pt}[dBm] - 10 \times \log_{10}(\mathrm {Pr}_{\mathrm {omni}}[\mathrm{mW}])
\end{equation}

\noindent where $\mathrm{Pr}(\theta_{r_w},\phi_{r_x},\theta_{t_y},\phi_{t_z})$ is the received power measured at one TX location and RX location combination for TX arbitrary pointing angles $\theta_{t_y}$ and $\phi_{t_z}$ in azimuth and elevation planes, respectively, and for RX arbitrary pointing angles $\theta_{r_w}$ and $\phi_{r_x}$ in azimuth and elevation planes, respectively. $y, z, w, x$ correspond to each rotation number of the TX and RX in the azimuth and elevation planes. $\mathrm {Pr_{\mathrm{omni}}}$ is the sum of all received power (with TX and RX antenna gain removed) from the sweeps in free space or foliage measurements, to recover the corresponding omnidirectional path loss $\PL_{\mathrm{omni}}$. $\mathrm{Pt}$ is the transmit power in dBm.

\begin{figure}
\centering
\includegraphics[width=3.5in]{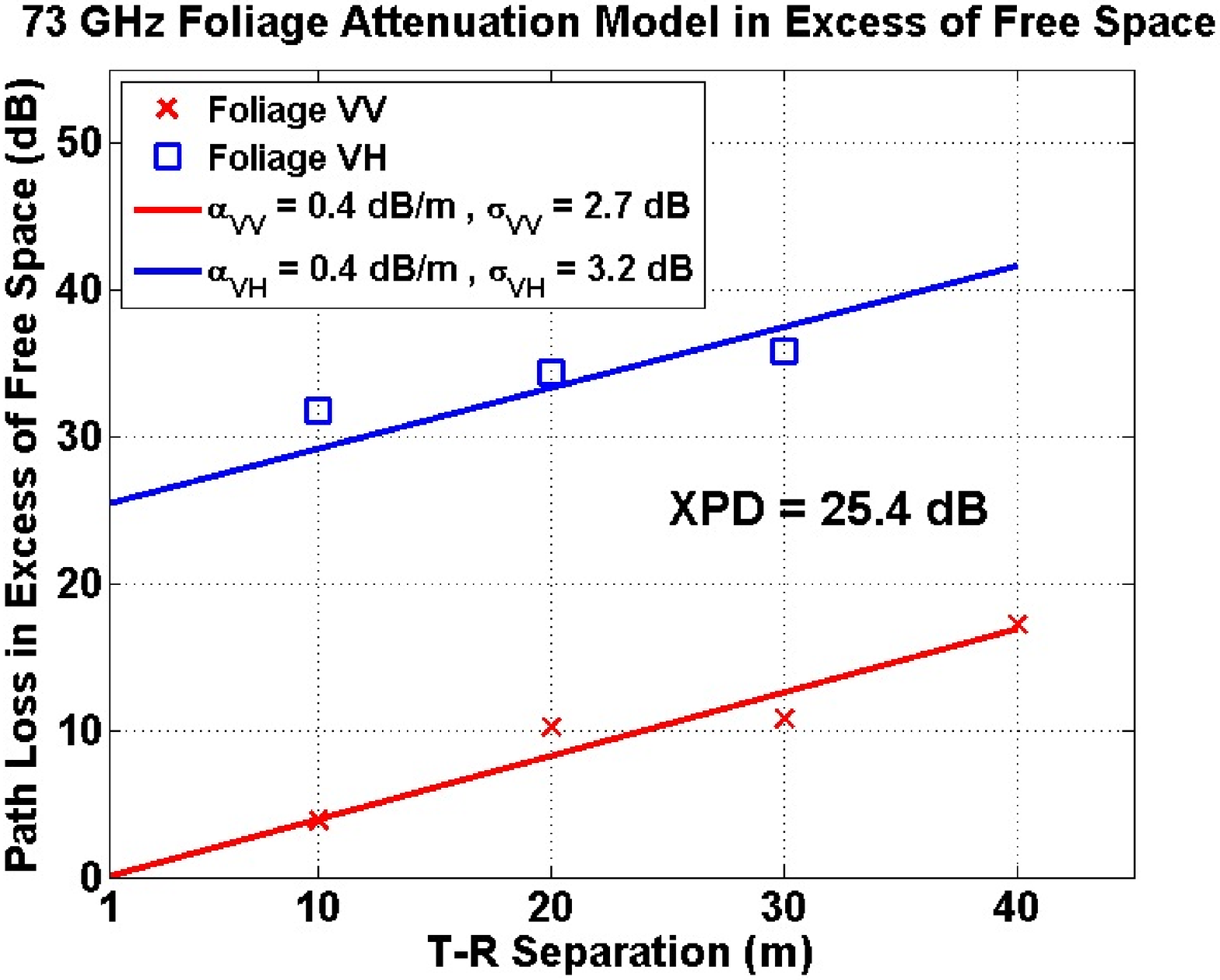}
\caption{73 GHz foliage attenuation path loss model in excess of free space for co- (V-V) and cross-polarized (V-H) antenna configurations for T-R separation distances of 10 m, 20 m, 30 m, and 40 m. The foliage type was common cherry tree branches, leaves, and trunks. }
\label{fig:Atten}
\end{figure}

The foliage attenuation rate $\alpha$ was obtained by applying the MMSE method over the omnidirectional path losses \cite{vogel1999effects,goldhirsh1998handbook}: 

\begin{equation}
\PL_{\mathrm {Foliage}}(d)[\dB]=\PL_{\mathrm {FS}}(d)+\alpha \cdot d[m]
\end{equation}
\begin{equation}
\alpha=\frac {  \sum_{i}^{}{[\PL_{\mathrm{Foliage}}(d_i)-\PL_{\mathrm{FS}}(d_i)  ] }  }   {\sum_{i}^{} d_{i}}
\end{equation}

\noindent where $\PL_{\mathrm{Foliage}}(d_i)$ and $\PL_{\mathrm{FS}}(d_i)$ are the omnidirectional path losses through foliage and in free space, respectively, measured at T-R separation distance $d_{i}$, as displayed in Table~\ref{tbl:PathLoss}. $\alpha$ (dB/m) represents the signal attenuation rate caused by foliage. Fig.~\ref{fig:Atten} shows the 73 GHz foliage attenuation path loss model for co- and cross-polarized antenna configurations. Note the received power for cross-polarization measurements at RX4 was too low to be detected by the channel sounder when the transmit power was set to -7.9 dBm. The foliage attenuation rates $\alpha$, were measured to be 0.4 dB/m with standard deviation of 2.7 dB and 0.4 dB/m with standard deviation of 3.2 dB for the co- and cross-polarized antenna configurations, respectively. The cross-polarization discrimination (XPD) factor was found to be 25.4 dB for free space measurements, and 26.8 dB for the foliage measurements, using 27 dBi horn antennas, which was extracted from the total path loss values. The high XPD for foliage measurements can be explained by the fact that the first foliage obstruction occurred after at least 5 m transmission in free space, the signal attenuated a lot for the first 5 m so that the de-polarization effect is not observed. The high XPD demonstrates the potential for orthogonal frequency reuse in foliage channels. The measured foliage attenuation rates can be used to estimate the total path loss through foliage in a common courtyard-style area in future mmWave ray-tracing algorithms and upper-layer system design.

\begin{table*}
\caption{ 73 GHz path loss values measured in free space ($\PL_{\mathrm{FS}} $ in $\dB$) and through foliage ($\PL_{\mathrm{Foliage}}$ in $\dB$), and corresponding foliage-induced path losses ($\Delta \PL$ = $\PL_{\mathrm{Foliage}} $- $\PL_{\mathrm{FS}} $) for co- (V-V) and cross-polarized (V-H) antenna configurations, at T-R separation distances of 10 m, 20 m, 30 m, and 40 m.}
 \label{tbl:PathLoss}
 \centering
 \begin{center}
  \begin{tabular}{|c|c|c|c|c|c|c|}
	
	\hline
	  \textbf{T-R Separation} & \multicolumn{3}{c|}{\textbf{V-V}}  & \multicolumn{3}{c|}{\textbf{V-H}} \\ \cline{2-7}
	 \textbf{Distance} & \bm{ $\PL_{\mathrm{FS}} $} & \bm{$\PL_{\mathrm{Foliage}}$} & \bm{$\Delta \PL$} & \bm{$\PL_{\mathrm{FS}}$} & \bm{$\PL_{\mathrm{Foliage}}$} & \bm{$\Delta \PL$} \\ 
	  \textbf{(m)} & \textbf{(dB)} & \textbf{(dB)} & \textbf{(dB)} & \textbf{(dB)} & \textbf{(dB)} & \textbf{(dB)} \\ \hline
	10   & 89.6  & 93.5   & 3.9  & 116.9 & 123.3  & 6.3   \\ \hline
	20   & 96.2  & 106.4  & 10.2 & 121.4 & 130.4  & 9.1   \\ \hline
	30   & 98.9  & 109.7  & 10.8 & 123.2 & 133.6  & 10.5   \\ \hline
	40   & 101.3 & 118.5  & 17.2 & 125.3 & -      &  -    \\ \hline

 \end{tabular}
\end{center}
\end{table*}

\subsection{Ground Reflection Coefficient}

Ground reflection measurements were conducted to compare the measured reflection coefficients computed based on different T-R separation distances and incident angles to the ground surface, with the theoretical Fresnel reflection coefficients \cite{GroundRef}. The ground reflection coefficients were recovered by determining the total free space distance between the TX and RX using knowledge of elevation angles and trigonometry, and removing the excess free space path loss and foliage attenuation from the total received power for both the co- and cross-polarized antenna configurations. The received power was obtained considering the initial specular reflection scenario (from M1 of the ground reflection measurements). The XPD of 26.8 dB from foliage measurements was removed from the total measured received power for the cross-polarized scenario, when recovering the ground reflection coefficients. The received power for free space and foliage measurements was calculated using \cite{Tedbook}: 

\begin{equation}
\mathrm \Pr_{\mathrm{FS}}=  \mathrm{P_{T}} \mathrm{G_{T}} \mathrm{G_{R}} \frac{\lambda^2 }{(4 \pi d_{i})^2}
\label{equ:PrFS}
\end{equation}

\begin{equation}
\mathrm \Pr_{\mathrm{Foliage}}=\mathrm{P_{T}} \mathrm{G_{T}} \mathrm{G_{R}} \frac{\lambda^2 }{(4 \pi d_{\mathrm{tot}})^2} \cdot 10^{\left(-\frac{d_{\mathrm{Foliage}} \cdot \alpha}{10}\right)} \cdot |\Gamma|^2
\label{equ:PrFo}
\end{equation}

\noindent where $d_{\mathrm{Foliage}}$ is the path length through tree canopies, $d_{\mathrm{tot}}$ is the total free space path length for the ground reflected ray, and $d_{i}$ is the horizontal distance between TX and RX (the vertical distance being negligible here). $\Pr_{\mathrm{FS}}$ and $\Pr_{\mathrm{Foliage}}$ are the omnidirectional total received powers (in mW) from the free space measurements and foliage ground reflection measurements, respectively.  $\alpha$ is the foliage attenuation rate in excess of free space ($\alpha=0.4$ dB/m for both V-V and V-H measurements). The magnitude of the measured reflection coefficient (in linear unit) was obtained by dividing Eq.~\ref{equ:PrFo} by Eq.~\ref{equ:PrFS} \cite{LandronRef}:

\begin{equation}
\label{Gamma}
|\Gamma|=\frac{d_{\mathrm{tot}}}{d_{i}} \cdot \sqrt{10^{\frac{d_{\mathrm{Foliage}} \cdot \alpha}{10}}} \cdot \sqrt{\frac{\mathrm{Pr_{Foliage}}}{\mathrm{Pr_{FS}}}}
\end{equation}

\begin{figure}
\centering
\includegraphics[width=3.5in]{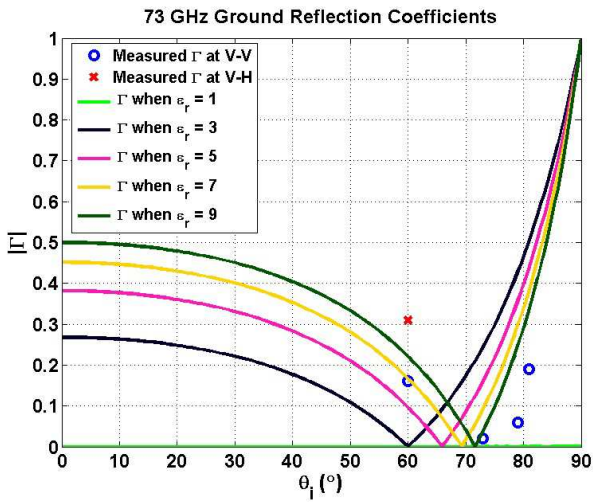}
\caption{73 GHz measured ground reflection coefficients ($|\Gamma|$) versus incident angles for V-V and V-H antenna polarization configurations using 27 dBi gain, 7$^{\circ}$ HPBW directional antennas. Theoretical reflection coefficient curves are shown for ground relative permittivity values ranging from 1 to 9 in increments of 2.}

\label{fig:Ref}
\end{figure}

All the distances and power levels are presented in Table~\ref{tbl:Pr}. Fresnel reflection coefficients are determined by material properties and incident angles \cite{Tedbook}. For the case when the first medium is free space and the permeability of the two media are the same, and when the electric field is parallel to the incident plane, the reflection coefficient can be simplified to \cite{LandronRef}:

\begin{equation}
\Gamma_{\parallel}'= \frac{\cos\theta_{t}-\sqrt{\varepsilon_{r}-\sin^2\theta_{i}}}{\cos\theta_{t}+\sqrt{\varepsilon_{r}-\sin^2\theta_{i}}}
\end{equation}

 \noindent where $\Gamma_{\parallel}'$ represents simplified parallel reflection coefficients when the electric field is parallel to the incident plane. $\theta_{i}$  is the angle of incidence, and $\theta_{t}$ is the angle of refraction, with respect to the normal incidence of the surface. $\varepsilon_{r}$ is the relative permittivity of the ground, consisting of soil, dirt, gravel, and fallen leaves. Measurements in \cite{gatesman2001physical} showed that the relative permittivity values of soil at 35 GHz and 94 GHz range from 3 to 5. Without the accurate values of permittivity of the ground material at 73 GHz, a series of typical values of soil permittivity are plotted in Fig.~\ref{fig:Ref}. The measured reflection coefficients (shown in Fig.~\ref{fig:Ref}) range from 0.02 to 0.31, for incident angles ranging from 60$^{\circ}$ to 81$^{\circ}$, for co- and cross-polarization antenna configurations, indicating 9.4 dB to 34 dB loss induced from ground reflection. Given the limited measured data set, especially due to the high attenuation in vertical-to-horizontal polarization antenna configuration, the measured result was not in good agreement with the theoretical value, future work will include more ground reflection measurements to validate the presented models. 

\begin{table*}
\caption{ Parameters used to calculate ground reflection coefficients according to Eq.~\ref{Gamma}. $d_{\mathrm{Foliage}}$ is the path length through tree canopies, $d_{\mathrm{tot}}$ is the total free space path length for the ground reflected ray, and $d_{i}$ is the horizontal distance between TX and RX (the vertical distance being negligible here). $\Pr_{\mathrm{FS}}$ and $\Pr_{\mathrm{Foliage}}$ are the omnidirectional total received powers (in dBm) from the free space and foliage ground reflection measurements, respectively. }
 \label{tbl:Pr}
 \centering
 \begin{center}
  \begin{tabular}{|c|c|c|c|c|c|c|}
	
	\hline

	  && & \multicolumn{2}{c|}{\textbf{V-V}}  & \multicolumn{2}{c|}{\textbf{V-H}} \\ \cline{4-7}
	   \bm{$d_{i}$}& \bm{$d_{\mathrm{tot}}$} & \bm{$d_{\mathrm{Foliage}}$} & \bm {$\Pr_{\mathrm{FS}}$}  & \bm{$\Pr_{\mathrm{Foliage}}$} & \bm{$\Pr_{\mathrm{FS}}$} & \bm{$\Pr_{\mathrm{Foliage}}$} \\ 
	   \textbf{(m)} & \textbf{(m)} & \textbf{(m)} & \textbf{(dB)} & \textbf{(dB)} & \textbf{(dB)} & \textbf{(dB)} \\ \hline
		10   & 11.7  & 8.0   & -44.2 & -64.6 & -72.4  &  -87.2  \\ \hline
		20   & 20.9  & 13.7  & -50.2 & -89.9 & -      &  -   \\ \hline
		30   & 30.6  & 21.0  & -53.1 & -86.1 & -      &  -   \\ \hline
		40   & 40.5  & 23.0  & -55.5 & -79.1 & -      &  -    \\ \hline

 \end{tabular}
\end{center}
\end{table*}

 \section{Conclusion }\label{sec:conclu}
 
This paper described mmWave outdoor foliage and ground reflection measurements at 73 GHz using a broadband sliding correlator channel sounder. The foliage attenuation rate was estimated to be 0.4 dB/m for both co- and cross-polarization antenna configurations with an XPD of 25.4 dB when using a pair of 27 dBi horn antennas. The high cross-polarization discrimination allows the use of orthogonal frequency reuse. These results can help in predicting path loss through foliage at 73 GHz for mmWave ray-tracing algorithms and system analyses. The measured ground reflection coefficients range from 0.02 to 0.31, or equivalently, 10.2 dB to 34 dB of reflection loss, indicating a reduced quality of the communication link and overall system throughput, which must be taken into account for wireless communication system design.

\bibliography{Foliage_Reference}

\end{document}